\newcommand{\ea}{{\it et al. }}
\newcommand{\mnras}{{\em Mon. Not. R. Astron. Soc. }}
\newcommand{\apj}{{\em Astrophys. J. }}
\newcommand{\aj}{{\em Astron. J. }}
\newcommand{\aaa}{{\em Astron. Astrophys. }}
\newcommand{\araa}{{\em Annu. Rev. Astron. Astrophys. }}
\newcommand{\apss}{{\em Astrophys. Space Sci. }}
\newcommand{\pasp}{{\em Publ. Astron. Soc. Pac. }}
\newcommand{\ptra}{{\it Phil. Trans. R. Soc. A}}

\documentclass{rspublic}

\begin{document}

\title[Binaries in star clusters]{Binaries in star clusters and the origin of the field stellar population}

\author[S.~P.~Goodwin]{Simon P.~Goodwin}

\affiliation{Department of Physics \& Astronomy, The University of Sheffield,
Hicks Building, Hounsfield Road, Sheffield S3 7RH, UK}

\label{firstpage}

\maketitle

\begin{abstract}{\bf binaries: general; stars: general; stars: formation;
galaxies: star clusters}Many, possibly most, stars form in binary and
higher-order multiple systems. Therefore, the properties and frequency
of binary systems provide strong clues to the star-formation process,
and constraints on star-formation models. However, the majority of
stars also form in star clusters in which the birth binary properties
and frequency can be altered rapidly by dynamical processing. Thus, we
almost never see the birth population, which makes it very difficult
to know if star formation (as traced by binaries, at least) is
universal, or if it depends on environment. In addition, the field
population consists of a mixture of systems from different clusters
which have all been processed in different ways.

\end{abstract}

\section{Introduction}

Observations suggest that a significant fraction of stars (perhaps
most) {\em in the field} are in binary or multiple
systems\footnote{Stars appear to have multiplicities from binaries to
septuples (e.g., Eggleton \& Tokovinin 2008). For brevity we shall use
`binary' to mean systems of any multiplicity, only drawing a
distinction when it is required.} (see, e.g., Duquennoy \& Mayor 1991;
Fischer \& Marcy 1992; Lada 2006; Eggleton \& Tokovinin 2008). It
seems to be impossible to dynamically produce binaries in anywhere
near the numbers observed, and so the vast majority of binaries must
have formed as binaries (Goodman \& Hut 1993).

As we have seen earlier in this volume (see Clarke 2010; de Grijs
2010; Lada 2010), a significant fraction of stars appear to form in
star clusters (see also Lada \& Lada 2003). Because of their high
densities, clusters are regions in which binary systems are likely to
be altered (either destroyed or changed) by dynamical
processes. Therefore, it is highly likely that far more binaries were
formed than are now observed, and even those binaries that survive may
have different properties at the present time compared to when they
formed.

The field is the sum of all star formation. As much of that star
formation was clustered, we expect that the field binary population
has undergone some (very probably significant) dynamical processing in
their birth clusters. Therefore, it is important to remember that the
field binary population is {\em not} the birth binary population. The
field is a mixture of systems from different environments, each of
which will have been processed to some degree.

The dynamical processing of binaries in clusters will also alter the
numbers and properties of various types of interesting astrophysical
systems such as blue stragglers, low- and high-mass X-ray binaries,
type Ia supernovae, and intermediate-mass black holes (Hut \ea
1992). However, in this review we will concentrate on `typical' star
formation that results in the bulk of the field, i.e., relatively
low-mass clusters (10--$10^5$ M$_\odot$) that disperse (are
destroyed?) within a few~Myr (see Lada \& Lada 2003). Thus, we will
ignore the extremely interesting, but relatively rare cases (at least
in the Galaxy) of extremely massive young clusters, or extremely
long-lived clusters in which many interesting dynamical processes
involving binaries occur.

In this contribution we will review our current understanding of two
key astrophysical problems: the universality of star formation and the
origin of the field. Binaries are an excellent tool with which to
attempt to answer these questions, as binary formation is a
fundamental and very common (possibly universal?) outcome of star
formation and so similarities and differences between binary
populations are indicators of the similarities and differences in the
star-formation process in different environments.

First, does star formation care about its environment (see also Clarke
2010; Lada 2010)? Is the outcome of star formation in cores of a
particular mass (at the end of the class 0/I phase) always
(statistically) the same? The initial mass functions (IMFs) of
different regions often appear very similar, but binary properties are
probably a far more detailed indicator of the similarity or otherwise
of star formation in different regions (Goodwin \& Kouwenhoven 2009).

Second, what is the origin of the field-star population? The field is
the sum of star formation in high- and low-mass clusters and isolated
star formation. Do we understand the origin of the field?

To attempt to answer these questions we will concentrate on local
regions and clusters for which we have detailed observations down to
low (often substellar) masses. Unfortunately, such clusters are of
low mass and often low density and the applicability of extending the
conclusions drawn from local star-forming regions to more extreme star
formation in massive young clusters and starbursts is debatable.

In \S2 we examine the observations of binary systems in the field and
in different clusters. In \S3 we discuss how binaries can be
dynamically processed in clusters and how this alters the birth binary
population. In \S4 we will investigate what we can infer about the
birth binary populations and if they are universal and discuss the
origin of the field binary population. We conclude in \S5.

\section{Observations of binaries}

Binary systems can be characterized by three fundamental
parameters. The first is the period or separation of the
system. Depending on how the binary has been observed, we may have
detailed orbital information (such as the semi-major axis and
eccentricity) or simply a projected separation. The second is the mass
of the primary star. And the third is the mass ratio, which gives the
relative masses of the components of the system, $q = M_2/M_1$, where
$M_1$ and $M_2$ are the masses of the primary and secondary stars,
respectively.

The fraction of stars which are in binaries is usually given by the
`multiplicity fraction' (often called just the `binary fraction')
\begin{equation}
mf = \frac{B + T + Q + ...}{S + B + T + Q + ...} ,
\end{equation}
where $S, B, T, Q$, etc., are the numbers of single, binary, triple,
quadruple, etc., {\em systems} (i.e., a binary system is made of two
stars). A number of other methods of quantifying binary fractions are
possible (in particular the `companion-star frequency'; see Reipurth
\& Zinnecker 1993 for a detailed discussion). 

Binaries are usually found in one of three ways: spectroscopic
binaries (from radial-velocity variations, biased to close
similar-mass companions), photometric binaries (from an `incorrect'
position on an Hertzsprung--Russell diagram, biased towards
similar-luminosity or mass companions) and visual binaries (stars too
close on the sky to be explained by chance projection, yet again
biased towards similar-luminosity or mass companions). Clearly, all
of these methods are biased, especially towards missing low-mass
(low-$q$) companions. It is worth keeping in mind throughout that
there could be hidden populations of binaries. A recent illustrative
example is the discovery of wide ($\sim 100$~au) brown-dwarf
companions to stars (e.g., Burgasser \ea 2005, 2007).

Given these biases, it is important in any survey of binaries to
understand the (primary) masses, separation ranges and mass-ratio
distributions that the survey is sensitive to. Often, comparisons of
different surveys are far more complex than they first appear (see,
for example, the careful comparison by Duch\^ene 1999).

\subsection{Binaries in the field}

The field provides the canonical binary properties to which other
observations are compared, in particular the field G-dwarf
distribution investigated by Duquennoy \& Mayor (1991; DM91). DM91
found that field G dwarfs have a multiplicity fraction of $\sim 0.6$,
a wide lognormal period/separation distribution with a peak at
approximately $10^4$~days/30~au and a mass-ratio distribution which
peaks at $q=0.2$ (but the latter depends on separation; see Mazeh
\ea 1992).

However, DM91 only probe a very small mass range around 1 M$_\odot$.
The vast majority of stars are M dwarfs, and the details of M-dwarf
multiplicity are far more unclear. Fischer \& Marcy (1992) and Reid
\& Gizis (1997) found a lower multiplicity fraction for M dwarfs, of 
$\sim 0.35$--0.4, with a roughly flat mass-ratio distribution (see
also Lada 2006 and references therein). Fischer \& Marcy (1992) found
that the separation distribution is similar to that of DM91 (although
it appears to peak at lower separations: see their figure~2).
Lada (2006) reviews recent M-dwarf multiplicity surveys and argues
that the M-dwarf multiplicity may be even lower, depending on the
binary fractions amongst very-low-mass stars.

The binary properties of brown dwarfs are somewhat unclear, but it
appears that their multiplicity fraction is very low, at 10--25\%
(e.g., Basri \& Reiners 2006; Law \ea 2007), and that they have a far
smaller range of separations, usually around 5--20~au (Close \ea 2003;
Basri \& Reiners 2006; Burgasser \ea 2007).

Binaries appear to be far more common in stars more massive than
1~M$_\odot$ than below, with a multiplicity fraction approaching 100\%
above a few M$_\odot$ (e.g., Abt 1983; Shatsky \& Tokovinin 2002;
Crowther \ea 2006; Kouwenhoven \ea 2007).

Many field stars are in higher-order multiples than simply binaries.
Eggleton \& Tokovinin (2008), in a survey of 4558 bright stars (almost
all $> 1$ M$_\odot$), find a raw (uncorrected for selection effects)
ratio of multiplicities of $2716:1438:285:86:20:11:2$ between one and
seven companions. This corresponds to at least 10\% of field systems
being higher-order multiples and Tokovinin \& Smekhov (2002) suggest
that this fraction could be 20--30\%. Surveys of young systems also
seem to find many higher-order systems (e.g., Leinert \ea 1993;
Koresko 2002; Brandeker \ea 2003; Correia \ea 2006; Lafreni\`ere \ea
2008; Connelley \ea 2008), but selections effects make solid estimates
of the higher-order multiplicity fraction difficult.

\subsection{Young binary systems}

In the last 20 years, there have been many studies of the binary
fractions of young (pre-main-sequence: PMS) stars. Note, however, that
many of these surveys have concentrated on visual, and therefore
relatively wide (hundreds of~au), binary systems.

Mathieu (1994 and references therein) summarized the binary fractions
and separation distributions of PMS stars and noted that there is a
significant excess of binaries with separations around 100~au compared
to the (G-dwarf) field (see also Patience \ea 2002). This has been
confirmed in many young star-forming regions, but it is clear that
different regions have different properties (see below).

Star-forming regions can be roughly divided into three categories:
isolated, low and high density. Their definitions are somewhat
arbitary (and can vary from author to author), but as a rough guide,
isolated star formation has stellar densities similar to the Galactic
field, of only a few stars~pc${-3}$. Low-density star-forming regions
(low-density clusters or associations) tend to have densities of
10--100 stars~pc${-3}$ (e.g., Taurus), while high-density regions are
more like the archetypal `cluster', with densities of $10^3$--$10^5$
stars~pc${-3}$ (e.g., from Orion to Westerlund~1).

Surveys of low-density star-forming regions tend to find an excess
binary fraction of a factor of 1.5--2 over the (G-dwarf) field value.
Leinert \ea (1993) and Ghez \ea (1993) found a significant excess of
binaries in Taurus, with almost everything $>0.3$ M$_\odot$ in a
binary system. Ghez \ea (1997) find twice the field binary fractions
in the star-forming clouds Chamaeleon (Cham), Lupus and Corona
Australis (see also K\"ohler \ea 2008).  K\"ohler \ea (2000) find an
excess of binaries by a factor of 1.6 in Scorpius (Sco)--Centaurus, as
do Patience \ea (2002) in $\alpha$ Perseus and Praesepe. Duch\^ene \ea
(2004, 2007) and Haisch \ea (2004) also find significant excesses of
binaries in very young (flat-spectrum and class I) sources in a number
of low-density regions (also found by Connelley \ea 2008, but see
below). Kraus \& Hillenbrand (2007) find an excess of wide
(300--1650~au) binaries in Taurus and Cham I, especially at higher
($>1$ M$_\odot$) masses. Lafreni\`ere \ea (2008) again find a
significant excess of binaries (and especially triples) in Cham I.

However, studies of higher-density star-forming regions tend to find
binary fractions similar to the field. Reipurth \& Zinnecker (1993)
found that PMS stars in groups of less than ten are twice as likely to
have a companion than those in groups with more than ten members. In
particular, the Orion Nebula Cluster (ONC) has a binary fraction
similar to that of the field (e.g., Petr \ea 1998; Kohler \ea 2006;
Reipurth \ea 2007), as do IC 348 (Duch\^ene \ea 1999) and $\eta$ Cham
(Brandeker \ea 2006). Connelley \ea (2008) also find that wide
(500--4500~au) class I binaries are less common in denser regions (at
odds with Duch\^ene \ea 2007, who find no environmental dependence).

There are a number of other observations that are worth noting.
Studies of very-low-mass objects (brown dwarfs and the smaller M
dwarfs) tend to find little or no evolution in the binary fraction
between even high-density regions and the field (e.g., Ahmic \ea
2007). However, Bouy \ea (2006) find evidence for a wide (100--150~au)
low-mass binary population in Upper Sco (USco), as do Konopacky \ea
(2007) in Taurus, which are not seen in large numbers in the field.

Kraus \& Hillenbrand (2007) find that wide binaries (330--1650~au)
show a strong mass--multiplicity relationship in Taurus, Cham I and
USco A, with these three clusters matching the difference between low-
and high-density regions well (Taurus and Cham I have an excess while
USco A looks like the field). But USco B is very strange, exhibiting
an excess of wide companions compared to the field at all masses,
possibly greatest at low masses, where approximately $35 \pm 15$\% of
M dwarfs have a wide companion. K\"ohler \ea (2000) also found that
the typical separation in USco B was about 30 times greater than in
USco A. This may be related to a possible very-low-mass binary
population with separations of 100--150~au in USco (Bouy \ea 2006).

Connelley \ea (2008) also find that the binary fractions $>1000$~au
appear to evolve during the class I phase, even in low-density
environments. They argue that this is the result of the decay of
higher-order systems (see \S3).

\section{Dynamical processing of binaries}

A key element in binary evolution in clusters is the dynamical
modification of binaries through encounters with single stars and
other binary systems (see also Vesperini 2010). This may result in a
change in orbital parameters if the encounter is relatively weak or in
the destruction (ionization) of one or the other binary system if the
encounter is strong. In addition, a binary may swap a single star or
component of the other binary for one of its components.

Heggie (1975) and Hills (1975) published seminal papers on dynamical
processing of binary systems (for a more gentle introduction to some
of these ideas, see the relevant sections of Binney \& Tremaine 1987;
see also Hut \ea 1992).

Binaries can be divided into three categories according to their
binding energies relative to their environment. `Hard' binaries are
very strongly bound and are unlikely to suffer disruptive encounters.
`Soft' binaries are very weakly bound and tend to be destroyed by an
encounter. `Intermediate' binaries lie between hard and soft, and can
sometimes be destroyed or significantly altered, but sometimes not
(these are clearly the most interesting category, but their study
requires $N$-body simulations). The evolution of binaries can be
summarized by the Heggie--Hills law: hard binaries get harder, while
soft binaries get softer with time.

The binding energy, $E$, of a binary with two components of mass $M_1$
and $M_2$ and semi-major axis $a$ is given by $E = -G M_1 M_2/2a$. If
the binary is located in an environment in which the average mass of a
star is $m$ and the velocity dispersion $\sigma$, then a binary is
hard if $|E|/m\sigma^2 \gg 1$, and soft if $|E|/m\sigma^2 \ll 1$ (and
intermediate if $|E|/m\sigma^2 \sim 1$).

It is important to remember that it is not just the hardness or
softness of a binary that is important in understanding whether that
binary will survive: the encounter rate also plays a vital role. Even
a soft binary may survive for a long time in regions where the
encounter timescale is very long. Therefore, the environment in which
a binary is found is of crucial importance. For example, the
hard--soft boundary for a 1 M$_\odot$/1 M$_\odot$ binary in the field,
which has a velocity dispersion of several tens of km s$^{-1}$, is a
few~au, but the vast majority of binaries are wider than this and
perfectly stable because the encounter timescale is many tens
of~Gyr. However, in a `typical' cluster with a velocity dispersion of
a few~km s$^{-1}$, the hard--soft boundary is tens of~au, but the
encounter timescale is only a few~Myr, resulting in rapid dynamical
processing.

It is also important to note that it is the {\it maximum} density a
cluster has had, rather than the current density, which is important
in setting the maximum size of binaries which we see in a cluster.
For example, Parker \ea (2009) suggested that to explain the binary
population of the ONC it must have been significantly denser in the
past and it was during this short-lived dense phase that the binary
properties were set (see also Kroupa \ea 2001; Scally \ea 2005; Moraux
\ea 2007; Bastian \ea 2008; Allison \ea 2009).

There are several main effects of encounters between binary systems
and single stars or other binary systems. Strong encounters can
destroy or heavily modify (e.g., cause a swapping of partners in)
binary systems, while even weak encounters can destroy soft systems or
change their orbital parameters (e.g., hardening or softening the
system, changing the eccentricity or inclination).

Higher-order systems (triples or higher) are often unstable and decay,
usually ejecting the lowest-mass member (Anosova 1986), as suggested
by Connelley \ea (2008) to explain the change of binary properties
with age in class I systems (see also Delgado Donate \ea 2004; Goodwin
\ea 2004; Goodwin \& Kroupa 2005).

In addition, internal evolution may play a role: the companion may
interact with the disc, migrate inwards or outwards, or magnetically
brake (see Kroupa 1995$b$; and references therein). However, these
processes probably only affect the tightest binaries and do so on a
timescale that is short compared to the dynamical timescale of a
cluster and so can be considered part of the `star-formation process'
(i.e., processes that occur in the class 0/I phase).

\section{The initial properties of binaries and the origin of the field}

As we have seen, there are a number of ways in which the birth
properties of binaries can be altered. In all but the loosest
associations and isolated star-forming regions we would expect some
dynamical processing by interactions between systems. And even in
isolated star formation we would expect decay of higher-order systems
or internal evolution to play a role. Thus, in {\em any} star-forming
region we can be almost certain that we are not observing the birth
population. And how the birth population has been altered will depend
on both the birth population and the local environment and its
evolution.

This makes answering the two questions with which we started
particularly difficult:

\noindent (1) What are the birth properties of binaries? Do they depend on
environment or are they universal?

\noindent (2) What is the origin of the field? This may be rephrased 
as: what is the sum of all the processing in all clusters of all
(different?) birth populations?

In short, the answer to (1) is that we are not certain, but what we do
apparently know is very confusing when applied to attempting to answer
(2). Such is science.

\subsection{The birth properties of binaries}

In \S2$\,b$ we reviewed a number of observations of young binary
systems. Two key points are obvious from the observations:

\noindent (1) Young stars tend to have an excess of binary companions
by a factor of 1.5 to 2 over the field.

\noindent (2) Denser star-forming regions tend to look like the field
and have few binaries with separations greater than a few thousand~au.

These two points can make sense within the context of clustered star
formation. Clusters will process binaries and denser clusters will
process them more efficiently. Therefore, the conclusion could be
drawn that most stars form as binaries with an excess at fairly wide
separations (hundreds to thousands of~au) and dense clusters will
rapidly process this initial population to look like the field.
Therefore, most stars form in multiples, many of which are dynamically
processed (e.g., Larson 1972, 2002; Mathieu 1994; Kroupa 1995$a,b$;
Goodwin \& Kroupa 2005; Goodwin \ea 2007).

However, while this scenario is almost certainly correct to some
(possibly great) extent, it is not clear if (a) most stars form as
binaries or (b) all star-forming regions produce the same population
which is then processed to produce different (field or cluster)
populations.

\subsubsection{Do most stars form as binaries?}

As pointed out by Lada (2006), most stars (90\%) are M dwarfs and most
M dwarfs are single. The exact importance of this depends on how
binarity is counted, if one third of M dwarfs form in binaries with
other M dwarfs then although two thirds of M-dwarf {\em systems} form
single, half of all M-dwarf {\em stars} form in binaries. However,
from the point of view of star formation, if two thirds of low-mass
{\em cores} form single stars, it would be the major mode of star
formation.

The importance of binary- versus single-star formation, however,
depends on what fraction of the initial M-dwarf binary population is
dynamically destroyed. This, in turn, depends on the initial
separation distribution of low-mass stars. There is some evidence for
a wide (100--150~au) low-mass binary population in low-density regions
(Bouy \ea 2006; Konopacky \ea 2007) which, if common, would be
expected to be very susceptible to dynamical destruction (Goodwin \&
Whitworth 2007). To make binary formation the major mode of star
formation, only around 20\% of the birth population of M dwarfs would
have to be in wide binaries.\footnote{For example, from 100 systems if
70 M dwarfs form as binaries, of which 20 are wide binaries, and 30
form as singles, then the destruction of the wide binaries would
produce 50 (of 120) binary systems and 70 (of 120) single stars (as
each binary destruction would produce two single M dwarfs).}

From theory, it might be expected that many low-mass cores only form
single stars. If disc fragmentation is the most common mode of binary
formation (see Goodwin \ea 2007), then very-low-mass cores should not
form binary systems. The minimum mass for fragmentation in a disc is
probably a few Jupiter masses (say, 0.005 M$_\odot$; Whitworth \&
Stamatellos 2006). For a disc to fragment, its mass must be
significantly greater than this minimum mass to collect enough
material to fragment without being sheared apart. Therefore, discs
that fragment must probably be $>0.1$ M$_\odot$ during the earliest
phases of star formation. Most of this disc material will accrete onto
a component of the binary, resulting in a total system mass of $>0.1$
M$_\odot$ (in addition to the material that was already in the
primary), suggesting a minimum system mass for disc fragmentation as a
mode of binary formation of perhaps 0.2 M$_\odot$, i.e., mid-M dwarfs
(interestingly close to the point at which Maxted \ea 2008 find a
dearth of low-mass binaries). However, at least some wide, low-mass
binaries do exist, possibly a significant number, which present
problems for star-formation models.

On the other hand, to produce the IMF from observed core-mass
functions, the efficiency of turning cores into gas must be only
around 30\% (Alves \ea 2007; Goodwin \ea 2008). Therefore, 70\% of
the gas initially in a core must not be accreted onto the stars (why?
how?), so possibly much of the material in the disc may not end up on
the stars.

In summary, if most low-mass stars form single, then most stars form
single. However, it is unclear what the birth separation distribution
of low-mass stars is (i.e., are the wide, low-mass populations common
or rare at birth?) and without this knowledge it is impossible to
assess the degree of dynamical processing of low-mass birth binaries.

\subsubsection{Is star formation universal?}

Do all stars form the same way? Do cores of a particular mass always
produce the same (statistical) outcome, or does this depend on
environment? Can the differences between loose associations be
explained as the outcome of different levels of processing of the same
birth populations?

The simplest null hypothesis is that all star formation is the same in
all environments and is then dynamically modified to produce different
populations in different environments (Kroupa 1995$a,b$). This
approach is very successful on a number of counts. The lack of wide
binaries ($>1000$~au) in dense clusters (like the ONC) compared to
loose associations and the field (Scally \ea 1999) is explained by the
almost complete dynamical destruction of such binaries. The
underabundance of intermediate-separation (few hundred~au) binaries in
clusters and the field compared to T Tauri stars (see \S2$\,b$) is
explained by the partial destruction of such binaries in clusters.
Indeed, Kroupa (1995$a$) managed to construct a birth binary
population that would produce the field when processed by a `typical'
cluster. There are, however, a number of problems with attempting to
produce a universal birth binary population.

Importantly, it is unclear what a `typical' cluster is. Clusters
appear to form with a mass function $\propto M_\mathrm{cl}^{-2}$ (Lada
\& Lada 2006), which would imply that an equal mass of stars form in
equal logarithmic mass bins. Therefore, $10^5$ M$_\odot$ clusters
produce as many stars as $10^2$ M$_\odot$ clusters.

Also, different star-forming regions appear to form different birth
populations. K\"ohler \ea (2000) and Kraus \& Hillenbrand (2007) find
significant differences between the apparently similar USco A and B
associations, with USco B having significantly wider binaries,
especially at low masses (see also Bouy \ea 2006). Kraus \&
Hillenbrand (2007) discuss the differences between USco A and B and
argue that USco B is a lower-mass association (similar to Taurus or
$\rho$ Ophiuchus) and possibly not associated with the more massive
Sco OB association.

USco B has probably always been a low-mass association, given its wide
binary population. However, Kraus \& Hillenbrand (2007) show that
USco B is different from the low-mass associations Taurus and Cham I in
that it has far more wide (300--1650~au) low-mass binaries than either
of these associations (around 30\% of 0.1--0.25 M$_\odot$ systems
compared to only a few~percent in Taurus and Cham I). Did such
systems form in Taurus and Cham I only to be broken up? (Such major
dynamical evolution seems unlikely unless Taurus and Cham I were
significantly denser in the past.)

In addition, to explain the differences between USco A and B from
purely dynamical processing, USco A must have been significantly
denser in the past. Preibisch \ea (2002) suggest that the entire USco
region was initially large ($\sim 25$~pc) as it has a large size and
low velocity dispersion, which suggests that the entire region was
initially of low density. This analysis includes USco A and B as being
part of the same star-forming event, while Kraus \& Hillenbrand (2007)
suggest that USco B should be treated separately. Even with this
caveat, it suggests that USco A should not have undergone significant
dynamical processing, making the differences difficult to explain by
anything other than different birth populations.

It might be expected that (binary) star formation in different regions
is different. In dense clusters such as the ONC (which was potentially
significantly denser in the past; see above), the average distance
between stars is only a few thousand~au. At such densities, does it
make sense to consider binaries forming with separations approaching
the average separation? However, this depends on believing that dense
clusters like the ONC {\em formed} dense, which they may well not have
(Allison \ea 2009) and that low-density clusters such as Taurus formed
at low density rather than in relatively dense low-$N$ clumps (see,
e.g., Kroupa \& Bouvier 2003).

In summary, it is very unclear if binary star formation is
`universal'. Some observational evidence points to differences between
different star-forming regions that cannot be explained by dynamical
processing, as we believe that these regions are dynamically
young. Unfortunately, the exact form of these differences is difficult
to determine, in particular because some dynamical processing must
occur in even low-density regions (even if this is almost all internal
decay).

\subsection{The origin of the field binary population}

The field binary population is the sum of all binaries (and single
stars) released from all star-forming regions after their dissolution.
It is, therefore, the sum of both isolated and clustered star
formation and the binaries in clusters will have been dynamically
processed to at least some degree.

An important point to make at this point is that {\em the outcome of
star-formation theories/simulations should not and cannot be compared
directly to the field.} Even if the models are of isolated star
formation, and therefore dynamical processing is not important, in the
field, unprocessed and processed binaries are mixed.

As we have seen, star formation does not appear to produce a universal
birth population. It is unclear how and why binary properties vary
among star-forming environments (or even if `environment' covers a
single parameter such as density, or whether it is a complex mixture
of density, turbulence, magnetic field strength, chemistry or a host
of other variables; see Klessen \ea 2009). Within clustered
environments, the birth population is further dynamically modified in
a way that depends mainly on density (but the density can and does
change significantly on very short timescales).

Given this situation, it seems that attempting to derive the origin of
the field population is an impossible task. However, there are a
number of interesting constraints which we can apply to the field
population. In particular, we can construct a model of universal star
formation that explains the field population, differences between
clusters and the differences between the M- and G-dwarf binary
fractions.

Star-forming regions can be roughly divided into three groups
according to how much they will dynamically process their binaries.
High-density clusters (HDCs) will significantly process much of their
birth binary population. Low-density clusters (LDCs) will process
wide systems, but leave fairly close systems unaffected. Isolated
star formation (ISF) will probably only suffer decay and internal
evolution to modify their birth populations (how important is this?).

Between 75 and 90\% of stars form in clusters (Lada \& Lada 2003; Lada
2010), and the rest form as ISF. The initial cluster mass function is
roughly proportional to $M_\mathrm{cl}^{-2}$ (Lada \& Lada
2003). Clusters appear to form with masses between approximately
$10^1$ and $10^6$ M$_\odot$, so an equal mass of stars forms in
clusters $<10^{3.5} M_\odot$ as do above. If we take clusters with
masses below and above $10^{3.5}$ M$_\odot$ to be LDCs and HDCs,
respectively, then 40\% of stars form in HDCs, 40\% of stars form in
LDCs, and 20\% form as ISF (obviously, these numbers are very rough,
but they suffice for the following discussion).

Binaries can be divided into four groups according to their
separation, $a$, and so how they will be affected by dynamical
processing in these different environments (following Parker \ea
2009).

\noindent {\em Close binaries} ($a<50$~au) are unaffected by dynamical
processing in all but the most extreme environments. Around 50\% of G
dwarfs and 25\% of M dwarfs are in close binaries.

\noindent {\em Intermediate binaries} ($50<a/\mathrm{au}<1000$) are
processed to a significant degree in HDCs, and to a much lesser extent
in LDCs. Around 20\% of G dwarfs and 10\% of M dwarfs are in
intermediate binaries.

\noindent {\em Wide binaries} ($10^3<a/\mathrm{au}<10^4$) are almost
always destroyed in HDCs and are significantly processed in LDCs. They
can only survive in ISF. About 15\% of G dwarfs and 8\% of M dwarfs
are in wide binaries.

\noindent {\em Very wide binaries} ($a>10^4$~au) cannot survive in any
cluster. Indeed, it is difficult to see how they form in even ISF as
their separations are larger than the typical size of a core. It is
thought that the only way to make significant numbers of very wide
binaries may be during the destruction of clusters (M. B. N.
Kouwenhoven \ea, in prep.). If this is true, then no (or few) stars
actually {\em form} as very wide binaries. About 15\% of G dwarfs are
in very wide binaries, as are probably a few percent of M dwarfs.

Taking G dwarfs as an example, we can construct a universal birth
population which evolves to the observed field distribution. If we
assume that G dwarfs form as 30\% close, 15\% intermediate and 25\%
wide binaries, and 30\% single stars, then dynamical processing will
destroy a few percent of intermediate binaries (i.e., half of the
intermediate binaries in HDCs), most of the wide binaries (all in the
HDCs, half in LDCs). If wide binaries then form later (a big `if'),
then this will produce a field population with 30\% close, 12\%
intermediate and 10\% wide binaries, and nearly 50\% single stars, of
which a fifth must somehow form very wide binaries (taking the
single-star fraction from 50 to 40\%).

The assumption has been made that G dwarfs are not in binaries with
other G dwarfs so the destruction of a G-dwarf binary will not dilute
the G-dwarf binary fraction other than by creating a single G dwarf
where there was once a binary with a G-dwarf primary. This is probably
fairly reasonable for G dwarfs where only very-high-mass-ratio systems
are both G dwarfs. However, for M dwarfs, most are in binaries with
other M dwarfs and so every binary destruction will add two single M
dwarfs, rather than one. This would dilute the total M-dwarf binary
fraction to a lower value of only 40--45\%, only slightly higher than
observed.

This model also ignores the decay of higher-order multiple systems
(see above) which will dilute the binary fraction even further
(Goodwin \& Kroupa 2005), especially at low masses as ejected stars
will generally be of low mass (Anosova 1986; Reipurth \& Clark 2001).

Therefore, we have a model in which all stars of whatever mass form
with the same birth binary fractions and separation distributions,
which explains why (a) denser clusters look like the field, but with
few wide binaries, (b) low-density clusters have more wide binaries
and (c) there are more single M dwarfs than G dwarfs.

In summary, it is possible to construct a universal model of star
formation. However, this apparently contradicts the previous section
in which we saw that there appear to be different populations in
clusters which are difficult to explain by anything other than
different birth populations. Of course, star formation would never be
expected to be completely universal, but how common and how
significant are the differences between different regions? Also, can
our understanding of local, generally low-mass, cluster formation be
extended to more massive and extreme events at all, or are they
completely different again (see also de Grijs 2010)?

\section{Conclusions}

We initially asked two fundamental questions related to star
formation:

\noindent 1. Is star formation (as probed by binary properties, at
least) universal, or does it depend on environment?\\
\noindent 2. What is the origin of the field binary population?

As we have seen, the properties of binaries in different environments
{\em are} different. In particular, dense clusters have fewer binaries
and those that they have tend to be close or intermediate binaries
($<1000$~au). However, this difference can be explained by dynamical
processing of the initial binary population in a cluster.  If a
cluster such as the ONC did form a significant wide binary population,
it would have been destroyed by now. Low-density star-forming regions
may provide a clue to the birth properties of stars, but only if one
believes that star formation is universal. The difference between
Taurus and the ONC is striking, but can be explained by both a
different birth population or dynamical processing, or a mixture of
both (dynamical processing {\em must} have occured in the ONC). Of
particular interest in this regard are observations of USco,
especially the differences between USco A and B, both of which are
thought to have formed at low density but have very different binary
properties. This may indicate differences in the initial conditions of
the cloud. Could this be due to triggered star formation (see
Preibisch \ea 2002)?

It is crucial to remember that we will {\em never} see an intact birth
population, even in a young cluster. The only way to see a potentially
intact birth system is to examine the deeply embedded phases of star
formation. But even then, significant accretion and fragmentation to
form binaries is ongoing during the class 0 phase, so that we will see
`unfinished' systems. However, by the class I phase, dynamical decay
can have occurred. Does this mean that the `birth population' is a
meaningless phrase?

One of the few statements that we can make without argument is that
{\em the field is not the birth population}. The field is a mixture of
potentially different birth populations which have been processed to
different degrees in their birth clusters and then mixed.  Dynamical
processing destroys binary systems. Therefore, there {\em must} have
been more binaries formed than we see in the field (of course, if star
formation is not universal, then some regions may form like the field,
but not all).

Without an understanding of the universality or otherwise of star
formation in different environments, it is impossible to constrain the
origin of the field population. As we have seen, it is possible to
explain the field binary population as the result of a universal mode
of star formation that has been processed differently in
different-density environments. Equally, it is possible to explain it
as the sum of many different modes of star formation, each of which
was then processed in different ways (if this is the case, then we
have a vast parameter space of possible answers). For simplicity, we
would probably prefer that (binary) star formation is universal.
However, this might well not be the case.

\end{document}